\documentclass[12pt]{article}
\begin{document}
 
\begin{center}
{\bf{Comment on `` A simple model for DNA denaturation'' 
By T. Garel,  C. Monthus and H. Orland, Europhys. Lett. 55, 132(2001)'' }}\\
\vskip .5cm
\bf{\large{Somendra M. Bhattacharjee }}\\
{Institute of Physics, Bhubaneswar 751 005, India;
  email:somen@iopb.res.in }
\end{center}
  
Garel, Monthus and Orland\cite{gmo} (to be referred to as GMO) in a study
of DNA denaturation 
transition argued that the effect of mutual repulsion of the two
strands can be approximated by a long range interaction.
Such a replacement is  unjustified and can lead to disastrous
consequences. 

The  dimensionless Hamiltonian is
\begin{eqnarray}
  \label{eq:1}
  H&=& \sum_i^2   \int_0^N \! \!\!ds \,  \frac{1}{2} 
 \left (\frac{\partial {\bf r}_i(s)}{\partial s}\right )^{^2}
+\ \int_0^N \!\!\!\! ds \ V({\bf  r}_{12}(s))+H_{\rm ev},\\
\label{eq:2}
H_{\rm ev}&=& 
 u_0 \int_0^N \!\!\!\! ds \ ds' \delta_{\Lambda}({\bf
  r}_{1}(s) -{\bf r}_2(s')) \\
\label{eq:3}
&\approx& \int_0^N \!\!\!\! ds\
\frac{\alpha_d}{\mid {\bf r}_{12}(s)\mid^{d-2}},\quad {\rm( by \ GMO)}
\end{eqnarray}
where ${\bf r}_i(s)$ is the $d$-dimensional position coordinate of a
monomer point at a contour length $s$ of chain $i$, each of length
$N$, $V$ is the base pairing interaction at the same contour length
(``directed polymer'' interaction), $H_{\rm ev}$ is the mutual
excluded volume (ev) interaction represented by the $u_0>0$ term with
$\delta_{\Lambda}({\bf r})$ as the usual delta function with an
ultraviolet cut-off $\Lambda$ in the reciprocal space.  The
thermodynamic properties come from the partition function $Z=\int
{\cal D R} e^{-H}$, where the integration is over the configurations
of the polymers.  Eq. \ref{eq:3} is the replacement advocated by GMO,
obtained by a partial sum over the polymer configurations given by the
first term on the right hand side of Eq. \ref{eq:1}.  The base-pairing
interaction $V(r)$ has just been added on to this effective repulsion
in Eq. 16 of Ref. 1.

It is well-recognized that the self- and mutually avoiding chains can
be described by the prevalent renormalization group approach which
predicts that in the large length scale limit the chains are
described\cite{leibler} by the stable fixed point $u^*\sim
\epsilon\equiv 4-d$, where $u$ is an appropriate running dimensionless
variable.  From a dimensional analysis point of view,
$[u_0]=[\alpha_d]= L^{d-4}$ where $L$ is a length scale but there is a
major difference between the two terms in the renormalization group
(RG) framework\cite{kolo,smb1,smb2}.  While the $u_0$ term flows under
renormalization, thereby reaching a fixed point (fp) for $\epsilon>0$,
i.e., $d<4$, the singular $\alpha_d$ term {\it does not} get
renormalized because of the analyticity of the RG
transformation\cite{kolo}.  This tells us that for
$\epsilon\rightarrow 0$ the renormalized dimensionless mutual
avoidance, $u$, goes to zero (chains behave like noninteracting random
walkers up to log-corrections) but, on the contrary, $\alpha_4$ does
not vanish.  In technical terms, $\alpha_4$ remains {\it marginal}
while $u$ is a marginally {\it irrelevant} variable at $d=4$. In fact
$\alpha_4$ leads to continuously varying exponents as GMO also
rediscovered (see Eq. 33 of Ref. 1).  For $\epsilon>0$, $u$ reaches a
f.p. but $\alpha$ {\it does not}. Therefore, Eq. \ref{eq:3}, {\it the
  replacement proposed by GMO, leads to a major contradiction}, if we
do not want to discard RG.

Lastly, the effects of long-range interactions for directed polymers,
as done in Ref. \cite{gmo} are already available in Refs.
\cite{kolo,smb1,smb2}.

I thank ICTP for hospitality.

\end{document}